\documentclass[12pt]{article}

\usepackage{authblk}
\usepackage{booktabs}
\usepackage{float}
\usepackage{multirow}
\usepackage{array}
\usepackage{rotating}
\usepackage{caption}
\usepackage{graphicx,psfrag,epsf}
\usepackage{enumerate}
\usepackage[numbers]{natbib}
\usepackage{url}
\usepackage{bbm}
\usepackage{amsthm,amsmath,amssymb}
\usepackage{mathrsfs}
\usepackage{neuralnetwork, CJK}
\usepackage{calligra}
\usepackage{algorithm}
\usepackage{algorithmicx}
\usepackage{algpseudocode} 
\usepackage{amsmath} 
\usepackage{algorithm}
\usepackage{algpseudocode}
\usepackage[resetlabels,labeled]{multibib}
\usepackage[a4paper, margin=1in]{geometry}
\usepackage{setspace}
\usepackage{hyperref}
\usepackage{longtable}
\usepackage{adjustbox}
\hypersetup{
    colorlinks=true,       % 启用彩色超链接
    linkcolor=blue,        % 内部链接的颜色
    citecolor=blue,        % 引用文献的颜色
    filecolor=magenta,     % 文件链接的颜色
    urlcolor=cyan          % URL 链接的颜色
}

\title{Confidence interval estimation of mixed oil length with  conditional diffusion model}

\author[1]{Yanfeng Yang}
\author[1]{Lihong Zhang}
\author[1,*]{Ziqi Chen}
\author[1]{Miaomiao Yu}
\author[2]{Lei Chen}
\affil[1]{East China Normal University, No. 3663, North Zhongshan Road, Putuo District, Shanghai, 200062, China}
\affil[2]{China University of Petroleum (East China), No. 66, West Changjiang Road , Huangdao District, Qingdao, 266580, China}
\affil[*]{zqchen@fem.ecnu.edu.cn}

\begin{document}

\maketitle

\providecommand{\keywords}[1]
{
  \small	
  \textbf{\textit{Keywords: }} #1
}

\begin{abstract}
Accurately estimating the mixed oil length plays a big role in the economic benefit for oil pipeline network. While various proposed methods have tried to predict the mixed oil length, they often exhibit an extremely high probability (around 50\%) of underestimating it. This is attributed to their failure to consider the statistical variability inherent in the estimated length of mixed oil. To address such issues, we propose to use the conditional diffusion model to learn the distribution of the mixed oil length given pipeline features. Subsequently, we design a confidence interval estimation for the length of the mixed oil based on the pseudo-samples generated by the learned diffusion model. To our knowledge, we are the first to present an estimation scheme for confidence interval of the oil-mixing length that considers statistical variability, thereby reducing the possibility of underestimating it. When employing the upper bound of the interval as a reference for excluding the mixed oil, the probability of underestimation can be as minimal as 5\%, a substantial reduction compared to 50\%. Furthermore, utilizing the mean of the generated pseudo samples as the estimator for the mixed oil length enhances prediction accuracy by at least 10\% compared to commonly used methods. 
\end{abstract}


\section{Introduction}
\label{Introduction}
\begin{figure*}[thbp!]
    \centering
    \begin{minipage}[t]{1.0\linewidth}
        \centering
        \includegraphics[width=1\linewidth]{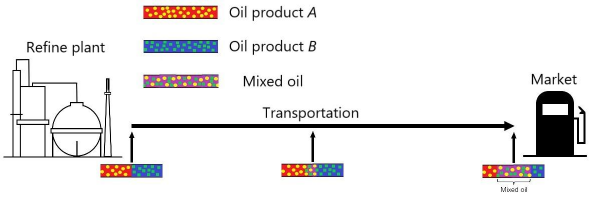}
    \end{minipage}

    \caption{The description of oil mixing phenomenon.}
    \label{fig:diagram of oil mixing phenomenon}
 \end{figure*}

With the advent of large-diameter and long-distance pipelines, the primary mode of transporting petroleum products has shifted to pipeline transportation \cite{trench2001pipelines,chen2021safety}, which is cost-effective, efficient, and safe \cite{huang2021carbon,mirhassani2013operational}.  
Various types of gasoline, kerosene, jet fuel, diesel, home heating oil, and liquefied gases like butane or propane \cite{mirhassani2011scheduling} are conveyed through a shared pipeline without any physical segregation between distinct batches \cite{abdellaoui2021multi}. 
As illustrated in Figure \ref{fig:diagram of oil mixing phenomenon}, when different batches move successively through a pipeline, a mixed oil section emerges between them \cite{austin1963mixing}, leading to the contamination of the pure products and a substantial compromise in their quality. This results in considerable economic losses due to the degradation in product quality \cite{du2023intelligent}. Therefore, the monitoring mixed oil interface is  essential  in multi-product pipelines. In particular, accurately calculating the mixed oil length is crucial  for optimizing the pipeline network's scheduling plan to guarantee products meet the required standards \cite{chen2021novel,patrachari2012conceptual}.

Austin and Palfrey \cite{austin1963mixing} proposed  two equations to calculate the length of mixed oil, with empirical parameters estimated from a large-scale industrial dataset. The  equations have found extensive practical use, even integrated into renowned commercial software for its convenience in rough estimation. However, its general applicability may be limited since the parameters of the above formula may vary in the different pipelines, environments, etc. \cite{li2023industrial}. Utilizing the  mechanism of convection and diffusion, Wang et al. \cite{wang2020study} found that considering the effect of near-wall adsorption can mitigate predictive deviations. By  considering the initial mixing, Hamidi \cite{hamidi2023numerical} introduced a two-dimensional numerical model and achieved more precise results. But implementing such sophisticated methods for real-time prediction of mixed oil length in long-distance industrial pipelines is challenging due to the substantial numerical burden involved.
In the fast-paced evolution of advanced optimization technology, researchers are delving into the utilization of machine learning algorithms. Their focus is on harnessing insights from historical data to construct models that can effectively calculate the length of mixed oil. Partial least squares and deep neural networks have been employed to develop predictive models for mixed oil length \cite{li2021data}. 
Chen et al. \cite{chen2021novel} integrated mechanistic models into the gradient boosting decision trees algorithm \cite{friedman2001greedy} to develop a dual-driver model for predicting the length of mixed oil. This integration resulted in enhanced prediction accuracy.
A serial-parallel residual-based machine learning algorithm, coupled with the Austin-Palfrey formula, has been applied to mitigate the impact of outliers in real pipeline mixed oil data \cite{yuan2021robust}.  
Utilizing the mechanism model, the Bayesian-Gaussian mixture regression algorithm and the Student's t mixture regression algorithm were employed to address the prediction of mixed oil length under variational pipeline conditions  \cite{yuan2023physics,yuan2023knowledge}.
Nevertheless, these methods only gave a point estimation, not an interval estimate; that is, the randomness of the sample leads to unrobust point estimation, while the estimate of the confidence interval takes this uncertainty into account.

In real-world scenarios,  despite assertions from existing methods about their accuracy in predicting mixed oil length,  these models frequently have a high probability (around 50\%) of underestimating the mixed oil length. This is primarily due to their failure to account for the statistical variability inherent in the estimated length of mixed oil. In Section \ref{Interval estimation of the length of the mixed oil}, experiments are carried out to substantiate this phenomenon. 
Furthermore, the accuracy measure does not differentiate between whether the predicted value is higher or lower than the actual value. To elaborate, consider two predicted values: the first predicted length is $s$ meters higher than the actual length, and the second predicted length is $s$ meters lower than the actual value. The RMSE value calculated using (\ref{rmse}) for these two predicted values remains identical.
When the predicted length exceeds the actual length of the mixed oil, workers can effectively remove the mixed oil from the pipeline. Conversely, if the predicted value falls below the actual length, complete exclusion of the mixed oil is not achievable, leading to increased economic losses.

The pioneering work on the diffusion model can be attributed to Jonathan Ho et al. \cite{ho2020denoising}. Renowned for its ability to accurately learn data distributions with limited samples, the diffusion model has found widespread applications across diverse domains
% Then diffusion model attributed to Jonathan Ho et al. \cite{ho2020denoising}  has found widespread applications across diverse domains
including computer vision \cite{dhariwal2021diffusion}, natural language processing \cite{li2022diffusion}, multimodal modeling \cite{rombach2022high, saharia2022photorealistic}, and time series processing \cite{rasul2021autoregressive}.
Further,  some improvements has been made to the original diffusion model, including advancements in sampling acceleration \cite{song2020denoising, bao2022analytic}, improvements in likelihood maximization \cite{nichol2021improved}, and innovations in conditional generation \cite{han2022card}. 
We learn the distribution of the mixed oil length given pipeline features using the diffusion model. Subsequently,  we generate  pseudo samples using the learned distribution and   use them to construct confidence interval estimation. Our approach takes into account the uncertainty in estimating the length of mixed oil. 
These intervals encompass the true length within a specified percentage $\alpha$, allowing workers to use the upper bound of the interval as a reference when excluding mixed oil.  The upper bound only underestimate the mixed oil length by a probability of $(1-\alpha)/2$, which can be much lower than $50\%$. Most importantly, in order to minimize the impact of the mixed oil segment on the quality of pure oil, on-site personnel typically allocate a certain margin during the batch cutting operations for oil \cite{li2017section}. However, this margin is usually determined solely based on experience, with significant interference from human factors \cite{harbert2008automation,zheng2023hybrid}. The interval estimation method proposed in this article provides crucial data support for determining the cutting margin. This is of significant importance for improving the operational efficiency of the refined oil distribution network.

We propose confidence interval estimation of mixed oil lengths based on the pseudo samples generated by the diffusion model to measure the uncertainty in mixed oil length predictions. Furthermore,
 we propose to predict the mixed oil length using the mean of the generated pseudo samples. 
 The main contributions of this paper can be summarized as:

\begin{enumerate}[1)]
 \item  Upon application to datasets collected from three operational pipelines,  we use the diffusion model to provide interval estimation for the mixed oil length, and the effectiveness of our interval estimation is confirmed.
The upper bound is suggested as a benchmark for the length of mixed oil when excluding it. It underestimates the mixed oil length with a probability of $(1-\alpha)/2$ theoretically, which can be much lower than $50\%$. In practical scenarios,  when $\alpha$ is set to $0.90$, the upper limit of the interval underestimates the mixed oil length  by only 6.8\% and the average radius of the interval is  approximately 300 meters.

\item Our proposed approach for predicting mixed oil length achieves a minimum of 14\% improvement in accuracy when compared to commonly employed methods.
    
\end{enumerate}

Given the presence of multiple notations in our paper, we provide detailed explanation of these symbols in Table \ref{tab:notation_maintext} and Table \ref{tab:notation_appendix}.

\begin{table}[thbp!]
\centering
\caption{The meanings of notations  in the main paper.}
\begin{adjustbox}{max width=\textwidth, }
\begin{tabular}{cc}
\hline
Notations & Meanings \\
\hline
$\alpha$ & The confidence level. \\
$L$ & The transportation distance. \\
$d$ & The inner diameter of the pipeline. \\
$Re$ & The Reynolds number. \\
$Re_j$ & The critical Reynolds number. \\
$C_0$ & The initial mixed oil length. \\
$L_e$ & The transportation distance in an equivalent pipeline corresponding to $C_0$. \\
$L_c$ & The calculated total transportation distance: $L+L_e$. \\
$C_{AP}$ & The calculated mixed oil length based on Austin-Palfrey equation. \\
$U(0,10)$ & The uniform distribution on $(0,10)$\\
$N(0,0.25)$ & A normal distribution with a mean of 0 and a variance of 0.25\\
$Y$ & A random variable, representing the mixed oil length. \\
$X$ & A random variable, representing the pipeline feature. \\
$E(Y|X)$ & The conditional mean of $Y$ conditioned on $X$. \\
$\{ x_i,y_i\}_{i=1}^n$ & The training dataset: $x_i$ is the pipeline feature, $y_i$ is the mixed oil length, and $n$ is the sample size. \\
$[y_i^{\alpha-low}, y_i^{\alpha-up}]$ & The confidence interval used for bounding $y_i$ at a probability of $\alpha$. \\ 
$P(y_i^{\alpha-low} \leq y_i \leq  y_i^{\alpha-up}|x_i)$
 & The probability of $y_i$ falling in $[y_i^{\alpha-low},y_i^{\alpha-up}]$ conditioned on $x_i$.\\
$Y|X$ & The distribution of $Y$ conditioned on $X$. \\
$N$ & The number of pseudo samples. \\
$\{ y_{i,l}\}_{l=1}^N$ & The pseudo samples corresponding to $y_i$. \\
$d_x$ & The dimension of the pipeline feature. \\
$\hat{Y}$ & An estimation of $E(Y|X)$. \\
$C_{AC}$ & The true value of the mixed oil length. \\
$t \in \{0,1,\cdots,T\}$ & The time steps in forward process and reverse process in the diffusion model. \\
$\{\beta_t\}_{t=1}^T$ & The diffusion schedule, i.e., pre-set parameters in the diffusion model. \\
\hline
\end{tabular}
\end{adjustbox}
\label{tab:notation_maintext}
\end{table}

\section{Methodology}
\label{Methodology}

In this section, we begin by presenting deterministic methods for predicting the length of mixed oil. Subsequently, we introduce our proposed approach to establish confidence interval estimator for the length of mixed oil.

\subsection{The deterministic methods}
\label{The deterministic methods}
Austin et al. \cite{austin1963mixing} formulated empirical equations for computing mixed oil based on actual pipeline data. Through extensive data analysis, they identified two distinct equations characterizing the relationship between mixed oil length and factors including transportation distance ($L$, in meters), pipeline inner diameter ($d$, in meters), and Reynolds numbers ($Re$).

Specifically, the critical Reynolds number \(Re_{j}\) is defined as:
\begin{equation}
\label{Re_j}
    Re_{j}=10000\exp(2.72d^{0.5}).
\end{equation}
Following  \cite{chen2021novel},  the initial mixed oil length $C_0$ can be regarded as the result of batch transportation in an 
equivalent pipeline with  length $L_e$: 
\begin{equation*}
\label{L_e}
\begin{aligned}
    L_e={\left(\frac{C_0Re^{0.1}}{11.75d^{0.5}}\right)}^2.
\end{aligned}
\end{equation*}
Then, the calculated total transportation distance is defined as 
\begin{equation*}
\label{L_c}
\begin{aligned}
    L_c=L+L_e.
\end{aligned}
\end{equation*}
%where \(d\) represents the pipeline inner diameter (in meters).
Subsequently, the Austin-Palfrey equations apply in two scenarios. When  Reynolds number \(Re\) is less than  \(Re_{j}\), the formula for calculating the mixed oil length is given by:
\begin{equation}
\label{C_AP_small_Reynolds}
    C_{AP}=18384d^{0.5}L_c^{0.5}Re^{-0.9}\exp(2.18d^{0.5}).
\end{equation}
%where \(L\) is the transportation distance (in meters). 
When \(Re\) exceeds \(Re_{j}\), the length of mixed oil can be significantly reduced \cite{austin1963mixing,chen2021novel}. The calculation of mixed oil length is expressed as follows:
\begin{equation}
\label{C_AP_big_Reynolds}
   C_{AP}=11.75d^{0.5}L_c^{0.5}Re^{-0.1}.
\end{equation}

In our datasets, the Reynolds number \(Re\) spans from \(1.124 \times 10^5\) to \(8.819 \times 10^5\). According to \eqref{Re_j}, \(Re_{j}\) ranges from \(3.514 \times 10^4\) to \(6.856 \times 10^4\). Therefore, the second Austin-Palfrey equation \eqref{C_AP_big_Reynolds} is applied to our real pipeline datasets.

\subsection{Confidence interval estimation}
\label{Confidence interval}

Numerous deterministic and machine learning methodologies \cite{austin1963mixing,yuan2021robust,chen2021novel,yuan2023physics} have been proposed to estimate the length of mixed oil based on pipeline features. Nevertheless, these methods typically provide only a single point estimate, neglecting the inherent statistical variability. The point estimation for mixed oil length tends to exhibit a substantial probability, approximately 50\%, of underestimating the actual length.
To address this limitation, the statistical concept of confidence interval estimation can be used. Confidence interval estimation provides not just a point estimate but also a range, with a lower and upper bound, effectively addressing the uncertainty associated with point estimation. 

\begin{figure*}[thbp!]
    \centering

    \begin{minipage}[t]{1.0\linewidth}
        \centering
        \includegraphics[width=0.8\linewidth]{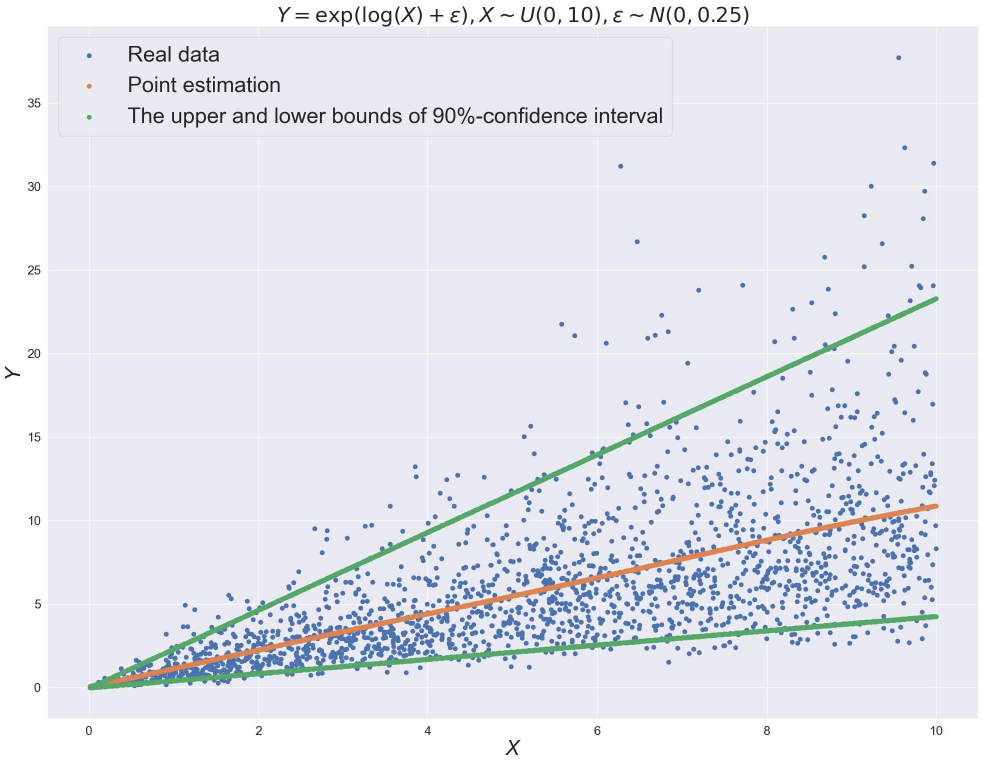}
    \end{minipage}
    
    \caption{Comparison of point estimation and confidence interval estimation. }
    \label{fig:comp_of_point_interval}
 \end{figure*}

Ignoring the  definition of $X$ and $Y$ in Table \ref{tab:notation_maintext} in this paragraph temporarily,  let $X\sim U(0,10)$ and generate $Y=\exp(\log(X)+\epsilon)$ with $\epsilon\sim N(0,0.25)$. An intuitive comparison of point estimation and confidence interval estimation is provided in Figure \ref{fig:comp_of_point_interval}. As depicted in the figure, the variance of $Y$ increases with $X$ due to the specific data generation mechanism, and the confidence interval can capture the uncertainty of $Y$ as $X$ varies.   In contrast, point estimation only reflects the expected value of $Y$ given $X$, $E(Y|X)$, and cannot capture the uncertainty of $Y$.

 %confidence interval has the capability to capture the variance of $Y$ as it varies with $X$. 

\begin{figure*}[htbp]   
  \centering            
  % \subfloat   
  {
      \includegraphics[width=0.6\textwidth]{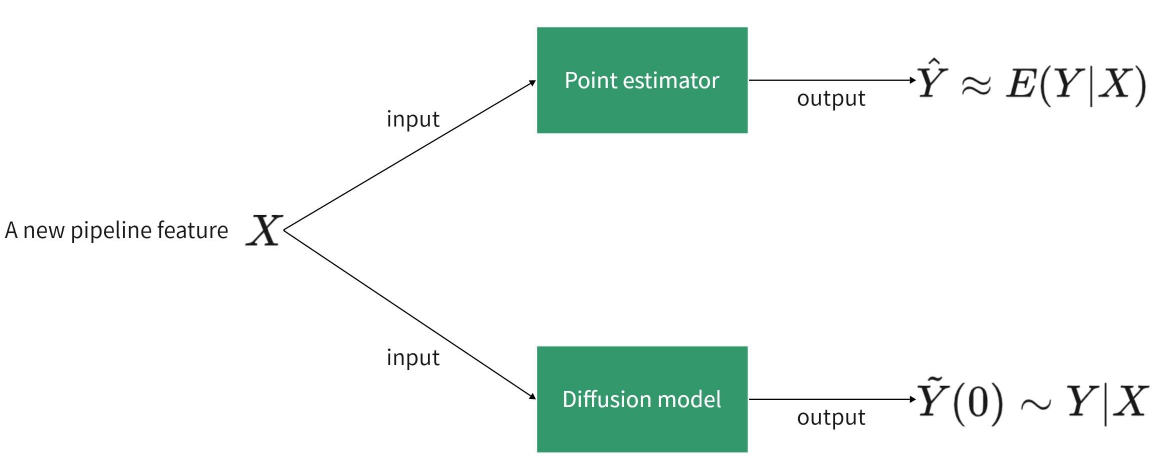}
      % 123456
  }

  {
      \includegraphics[width=0.45\textwidth]{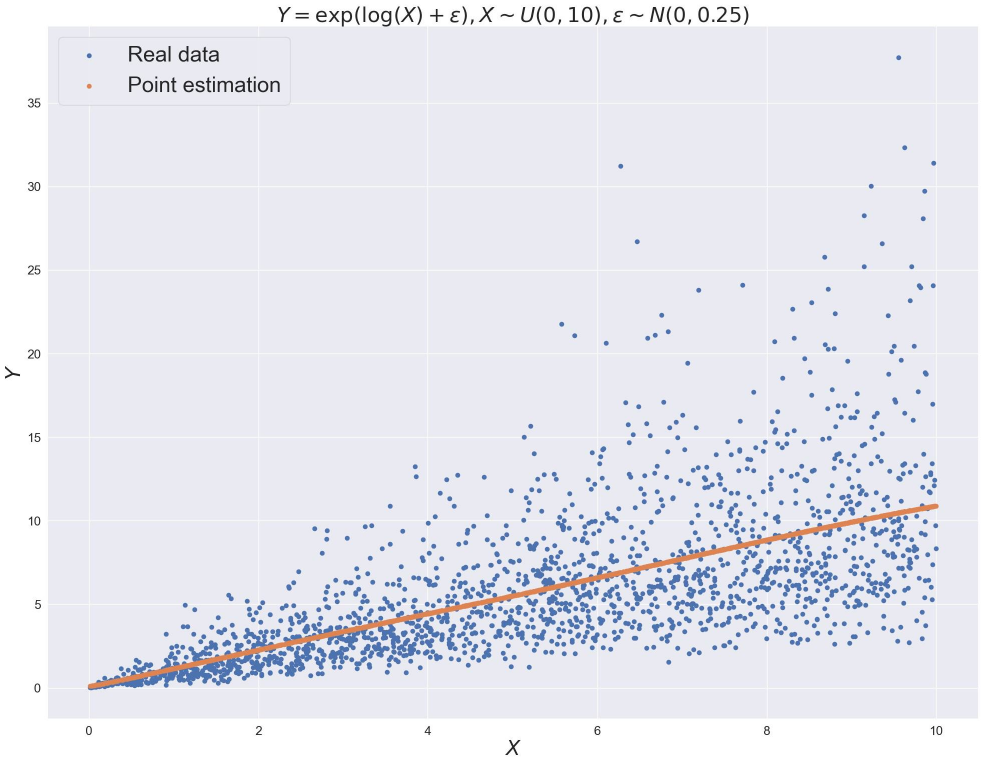}
      % 123456
  }
  % \subfloat
  {
      \includegraphics[width=0.45\textwidth]{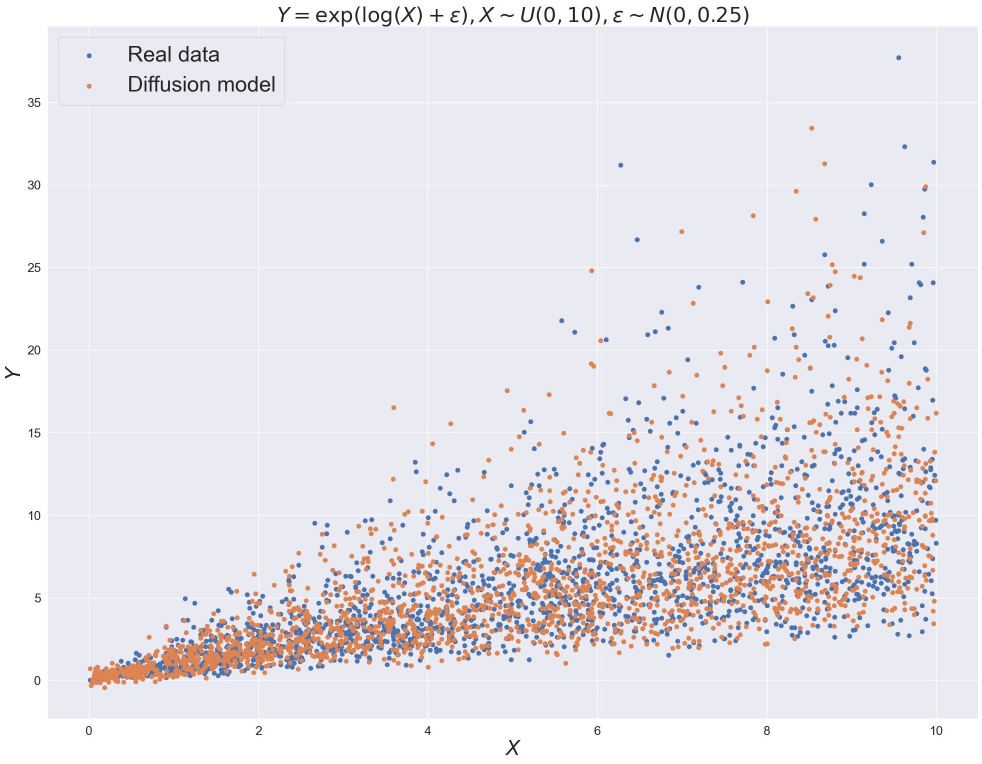}
  }
  \caption{Comparison of the outputs of point estimator and diffusion model.} 
  \label{fig:difference_point_and_diff}       
\end{figure*}

Next, we show the mathematical definition of confidence interval estimation. Considering a dataset encompassing $n$ samples characterized by the observed pipeline features $\{x_i \}_{i=1}^n$, our objective, at a given confidence level $\alpha$, is to construct a confidence interval for the unknown mixed oil length $y_i$. This interval, denoted as $[{y_i}^{\alpha-low},{y_i}^{\alpha-up}]$, is defined to satisfy 
\begin{equation*}
\label{eq:confidence interval}
    P(y_i^{\alpha-low} \leq y_i \leq y_i^{\alpha-up}|x_i) = \alpha,
\end{equation*} 
and is established based on the corresponding pipeline features $x_i$,  where $i$ ranges from 1 to $n$.
Assuming knowledge of the conditional distribution of $Y|X$ with $X$ representing the pipeline features and $Y$ representing the mixed oil length, we can generate $N$ pseudo samples denoted as $\{y_{i,l}\}_{l=1}^N$ for the given $x_i$. Subsequently, from these $N$ samples, we derive the upper $(1-\alpha)/2$ sample quantile and lower $(1-\alpha)/2$ sample quantile to obtain ${y_i}^{\alpha-up}$ and ${y_i}^{\alpha-low}$, respectively. In practical applications, workers can then exclude ${y_i}^{\alpha-up}$ meters of mixed oil, ensuring the exclusion of mixed oil by $(1+\alpha)/2$. This probability can significantly  exceed the probability (approximately 50\%) of excluding mixed oil based on a single-point estimation. It's worth noting that the confidence level $\alpha$ can typically be set to either $90\%$ or $95\%$.

To conclude, the advantages of confidence interval estimation compared to point estimation lie in several aspects:
\begin{enumerate}[1)]
  \item \textbf{Considering uncertainty:} Confidence interval estimation takes into account statistical error, thus better reflecting the true nature of the estimation results. In contrast, point estimation only provides a single estimator, inevitably lacking robustness and under-estimating the mixed oil length.
  \item \textbf{More reliable inference:} Confidence interval estimation provides a range that can be used to judge where the true value of mixed oil length. Such judgments are more reliable because they consider the uncertainty of the estimation, rather than relying solely on a point estimator.
  \item \textbf{Greater adaptability:} Confidence interval estimation allows for the adjustment of the confidence level based on specific needs to fulfill various inferential requirements.
  %Confidence interval estimation can adjust the confidence level according to demand to meet different inferential requirements. 
  This makes confidence interval estimation more flexible and adaptable.
 \end{enumerate}

\subsection{Diffusion model}
\label{Diffusion models}
In Section \ref{Confidence interval}, we assume that the conditional distribution of 
$Y$ given 
$X$ is known, and this serves as the foundation for constructing confidence interval estimates.
Nonetheless, in reality, we do not have precise knowledge of this conditional distribution. So, we propose to learn this distribution using the conditional diffusion model \cite{han2022card}. Specifically, let $\{x_i, y_i\}_{i=1}^{n}$ represent our training dataset, consisting of $n$ samples used to train the diffusion model. In this context, $x_i$ denotes the $d_x$-dimensional pipeline features, and $y_i$ represents the corresponding mixed oil length. Utilizing Algorithm \ref{algo:card}, the diffusion model has the capability to learn the conditional distribution $Y$ given $X$ from a given training set, and subsequently generate a predefined number of pseudo samples. Detailed descriptions of the diffusion model can be found in Appendix \ref{appendix:diffusion model}.

Within the framework depicted in Figure \ref{fig:comp_of_point_interval}, the difference between the  point estimator and the diffusion model   is illustrated in Figure \ref{fig:difference_point_and_diff}. Specifically, the point estimator  only learns the conditional mean $E(Y|X)$ and outputs an estimator of $E(Y|X)$, whereas the diffusion model learns the conditional distribution of $Y|X$ and  outputs pseudo samples  from $Y|X$.

\begin{algorithm*}[htb!]  %ht!参数是调整算法在文章中的位置
	\renewcommand{\algorithmicrequire}{\textbf{Input:}}
	\renewcommand{\algorithmicensure}{\textbf{Output:}}
	\caption{Train the conditional diffusion model and generate pseudo samples}  
	\label{algo:card}
	\begin{algorithmic}[1] %每行显示行号
		\Require Dataset $\{x_i, y_i\}_{i=1}^{n}$, max timestep $T$, diffusion schedule $\{\beta_t\}_{t=1}^T$, pre-trained model $f_\phi(X)$, an optimizer.
		\Ensure  A function to generate pseudo sample $\Tilde{Y}(0)$. 

            \Statex
            
            \State \textbf{Train the conditional diffusion model} 
            \State Let $\textbf{X}= (x_1,x_2,\cdots,x_n)^T, \textbf{Y}(0)=(y_1,y_2,\cdots,y_n)^T$
            \State Initialize a neural network $nn_{\theta}$, $\theta$ represents the parameters of $nn_{\theta}$
            \State \textbf{While} not converge:
            \State \quad Randomly draw $t$ from $\{1,2,\cdots,T\}$ 
            \State \quad Let $\alpha_t=1-\beta_t$,
            $\bar{\alpha}_t=\prod_{s=1}^{t}\alpha_s$
            \State \quad Draw $\epsilon_{0 \rightarrow t} \sim N(0,I_{n \times n})$
            \State \quad Let $\textbf{Y}(t)=\sqrt{\bar{\alpha}_t}\textbf{Y}(0)+ \left(1-\sqrt{\bar{\alpha}_t}\right)f_\phi(\textbf{X}) + \sqrt{1-\bar{\alpha}_t}\epsilon_{0 \rightarrow t}$
            
            \State \quad Compute $\textit{\textbf{L}}_{\theta}=\| \epsilon_{0 \rightarrow t} - {nn}_{\theta}\left(\textbf{X},\textbf{Y}(t), f_\phi(\textbf{X}), t\right)\|^2$ 
            \State \quad Take optimization step on $\textit{\textbf{L}}_{\theta}$ and update $\theta$ via the optimizer
            \State \textbf{end While}

            \Statex

            \State \textbf{Function} $G(X)$: \Comment{Function to generate pseudo samples based on each new pipeline feature $X$}
            \State \quad Draw $\Tilde{Y}(T) \sim N\left(f_{\phi}(X),1\right)$
            \State \quad \textbf{For} $t$ from $T$ to $1$:
            \State \quad \quad Let $\Hat{Y}^t(0) =\dfrac{1}{\sqrt{\bar{\alpha}_t}}\left[ \Tilde{Y}(t)-\left(1-\sqrt{\bar{\alpha}_t}\right)f_\phi(X)-\sqrt{1-\bar{\alpha}_t} nn_{\theta}\left(X,\Tilde{Y}(t), f_\phi(X), t \right)\right]$
            \State \quad \quad Let $\tilde{\mu}\left(\hat{Y}^t(0),\Tilde{Y}(t),X\right)=\dfrac{\beta_t \sqrt{\bar{\alpha}_{t-1}}}{1-\bar{\alpha}_{t}}\hat{Y}^t(0)+\dfrac{\left(1-\bar{\alpha}_{t-1}\right)\sqrt{\alpha_t}}{1-\bar{\alpha}_{t}}\Tilde{Y}(t)+\left( 1+\dfrac{\left(\sqrt{\bar{\alpha}_{t}}-1\right)\left(\sqrt{\alpha_t}+\sqrt{\bar{\alpha}_{t-1}}\right)}{1-\Bar{\alpha}_t}\right) f_\phi(X)$
            \State \quad \quad Draw $\epsilon_{t \rightarrow (t-1)}$ from $N(0,1)$
            \State \quad \quad $\Tilde{Y}(t-1)=\tilde{\mu}\left(\hat{Y}^t(0),\Tilde{Y}(t),X\right)+\sqrt{\tilde{\beta_t}}\epsilon_{t \rightarrow (t-1)}$  \Comment{The definition of $\tilde{\beta_t}$ can be found in (\ref{beta t})}
            \State \quad \textbf{end For}
            \State \quad \textbf{Return} $\Tilde{Y}(0)$
            \State \textbf{end Function}

	\end{algorithmic}
\end{algorithm*}

\subsection{Comparable distributions }
\label{A serious difficulty: Distribution shift}
In this article, we apply our method to a dataset obtained from the Supervisory Control and Data Acquisition (SCADA) system. This dataset comprises samples from three operational pipelines. For the training phase of our diffusion model, we use the 1781 samples from the first two pipelines. Subsequently, we assess the performance of our trained diffusion model using 528 samples from the third pipeline. To be specific, we utilize the pipeline features from the third pipeline as input to generate pseudo samples and establish confidence intervals for the mixed oil lengths.

\begin{figure*}[htbp]   
  \centering            
  % \subfloat   
  {
      \includegraphics[width=0.45\textwidth]{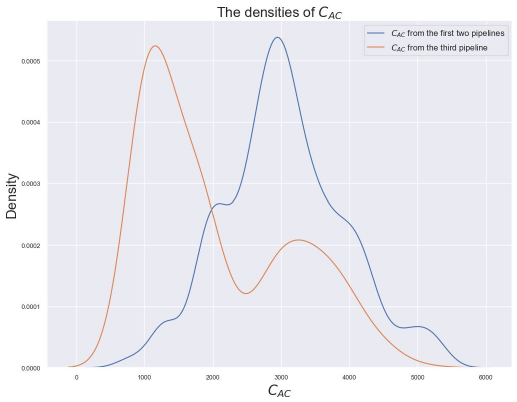}
      % 123456
  }
  % \subfloat
  {
      \includegraphics[width=0.45\textwidth]{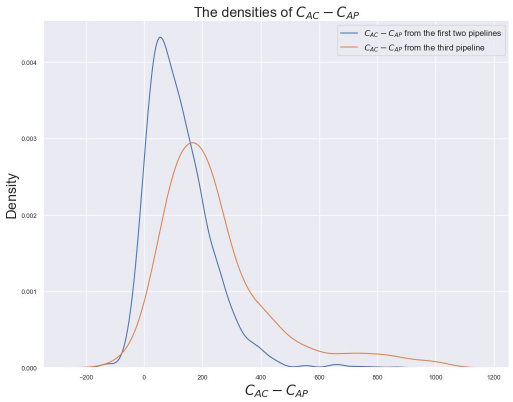}
  }
  \caption{Left panel: estimated density functions of $C_{AC}$ in the training and testing sets.
Right panel: estimated density functions of $C_{AC}-C_{AP}$ in the training and testing sets.} %It is noticeable that the density functions in the second graph are more similar to each other, indicating less severe distribution shift.}    
  \label{fig:density}       
\end{figure*}

Importantly, the diffusion model presented in Section \ref{Diffusion models} implicitly assume that the distribution of the mixed oil length for the first two pipelines (training dataset) matches or closely resembles the distribution for the third pipeline (testing dataset).  We use $C_{AC}$ to represent the true value of the mixed oil length. As depicted in the left panel of Figure \ref{fig:density}, significant disparities in the distribution of $C_{AC}$ exist between the first two pipelines and the third pipeline \cite{reynolds2009gaussian, yuan2021robust}. This difference  presents a considerable challenge to the generalization capability of our proposed diffusion model.

\begin{figure}[thbp!]
    \centering
    \begin{minipage}[t]{1.0\linewidth}
        \centering
        \includegraphics[width=0.6\linewidth]{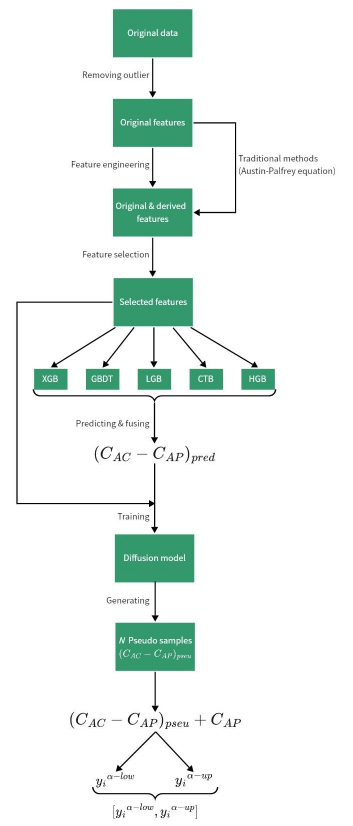}
    \end{minipage}

    \caption{The flow chart of our proposed method.} %}The definition of $C_{AC}$ and $C_{AP}$ can be found in Section \ref{Data processing} and Section \ref{The deterministic methods}.}
    \label{fig:flow chart}
 \end{figure}

We obtain a preliminary estimation of the mixed oil length $C_{AP}$ using the Austin-Palfrey formula mentioned in Section \ref{The deterministic methods}. Subsequently, we calculate the difference between the true value $C_{AC}$ and $C_{AP}$, i.e., $C_{AC}-C_{AP}$.
As shown in the right panel of Figure \ref{fig:density}, the distribution of $C_{AC}-C_{AP}$ in the first two pipelines is comparable to that in the third pipeline \cite{yuan2021robust}. Therefore,  in Section \ref{Diffusion models}, given the pipeline feature, instead of learning the distribution of the mixed oil length, we propose to learn the distribution of  $C_{AC}-C_{AP}$. 
For a pipeline feature from the third pipeline, once we generate $N$ pseudo samples of $C_{AC}-C_{AP}$ using the diffusion model, we can acquire $N$ pseudo samples of the mixed oil length by adding $C_{AP}$ to each sample within the pseudo samples of $C_{AC}-C_{AP}$. The flowchart depicting our proposed method can be found in Figure \ref{fig:flow chart}.

\section{Case Studies}
\label{Case Studies}

\subsection{Data processing}
\label{Data processing}
In this section, we apply our approach to a dataset gathered from the SCADA system. %Considering the presence of inevitable measurement errors, immeasurable noise, or errors in recording, there are outliers. Following \cite{chen2021novel},  we calculated the relative error between $C_{AP}$ and the actual value $C_{AC}$. If the relative error of a given sample exceeds $20\%$, it is identified as an outlier and removed from the dataset. Consequently, 26 samples from the first two pipelines are excluded from the dataset. 
Given the inevitable presence of measurement errors, immeasurable noise, or inaccuracies in recording, outliers are present in the dataset. Following the approach in \cite{chen2021novel}, we computed the relative error between $C_{AP}$ and the actual value $C_{AC}$. If the relative error of a specific sample exceeds $20\%$, it is designated as an outlier and subsequently removed from the dataset. Consequently, 26 samples from the first two pipelines were excluded.

\subsection{Feature selection}
\label{Feature Engineering}
The features from the collected on-site mixed oil data include transmission distance $L$, diameter $d$, initial mixed oil length $C_{0}$, and Reynolds number $Re$. %Before selecting the input features for the model, we exclude the outliers as described in Section \ref{Data processing}. 
 The second Austin-Palfrey equation \eqref{C_AP_big_Reynolds} is applied to our real pipeline datasets to calculate $C_{AP}$ as mentioned in subsection \ref{The deterministic methods}. We have conducted experiments  to demonstrate that the calculated value $C_{AP}$  is the most important feature. Motivated by \cite{chen2021novel}, the following two more features are as alternatives:
 \begin{equation*}
\label{d'}
\begin{aligned}
    d'=d^{0.5},~~~~~~Re'=Re^{-0.1}.
\end{aligned}
\end{equation*}
Based on the variable importance outputted by XGBoost \cite{chen2016xgboost}, we choose the initial oil mixture $C_{0}$, the formula-calculated value $C_{AP}$, and the transformed features $d'$ and $Re'$ as the input pipeline features for our proposed diffusion model.

\subsection{Pre-trained model and the diffusion schedule}% and modeling process}
\label{Parameter tuning and modeling process}

Five ensemble tree models, specifically XGBoost \cite{chen2016xgboost}, GBDT \cite{friedman2001greedy}, LGB \cite{ke2017lightgbm}, CTB \cite{prokhorenkova2018catboost}, and HGB \cite{hang2021gradient}, are utilized to capture the relationship between the selected pipeline features $X$ and $C_{AC}-C_{AP}$ individually. The pretrained model is obtained by combining the predictions from these five models. To be specific, the prediction of the pretrained model is the average of the results obtained from these five models.
 Within the diffusion model, the diffusion schedule $\{\beta_t\}_{t=1}^T$ is configured to increase linearly from $\beta_1=10^{-5}$ to $\beta_T=2 \cdot 10^{-3}$, with the maximum timestep $T$ set to 1000.

\subsection{Interval estimation of mixed oil length}
\label{Interval estimation of the length of the mixed oil}
\begin{figure*}[thbp!]
    \centering
    \begin{minipage}{1.0\linewidth}
        \centering
        \includegraphics[width=0.9\linewidth]{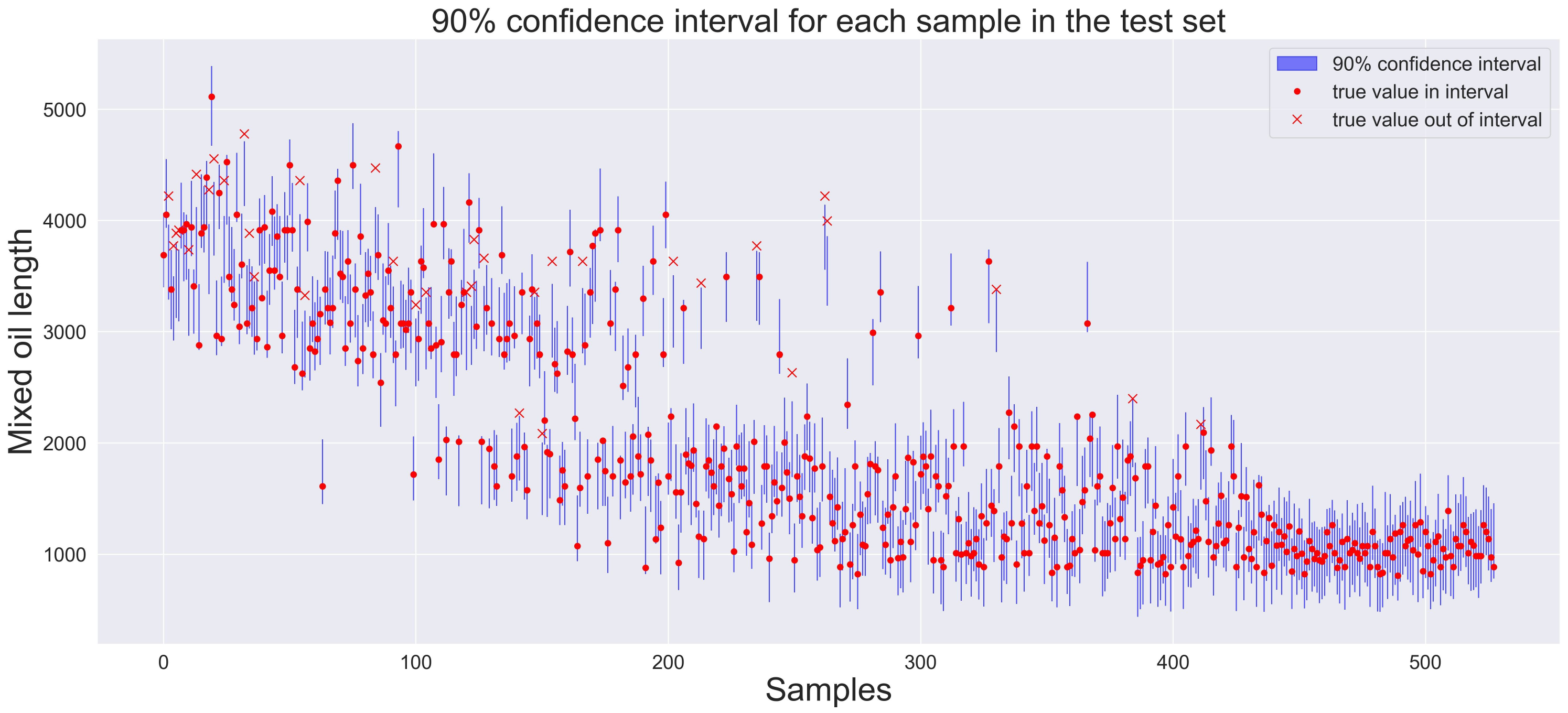} \\ \vspace{10mm}
        \includegraphics[width=0.9\linewidth]{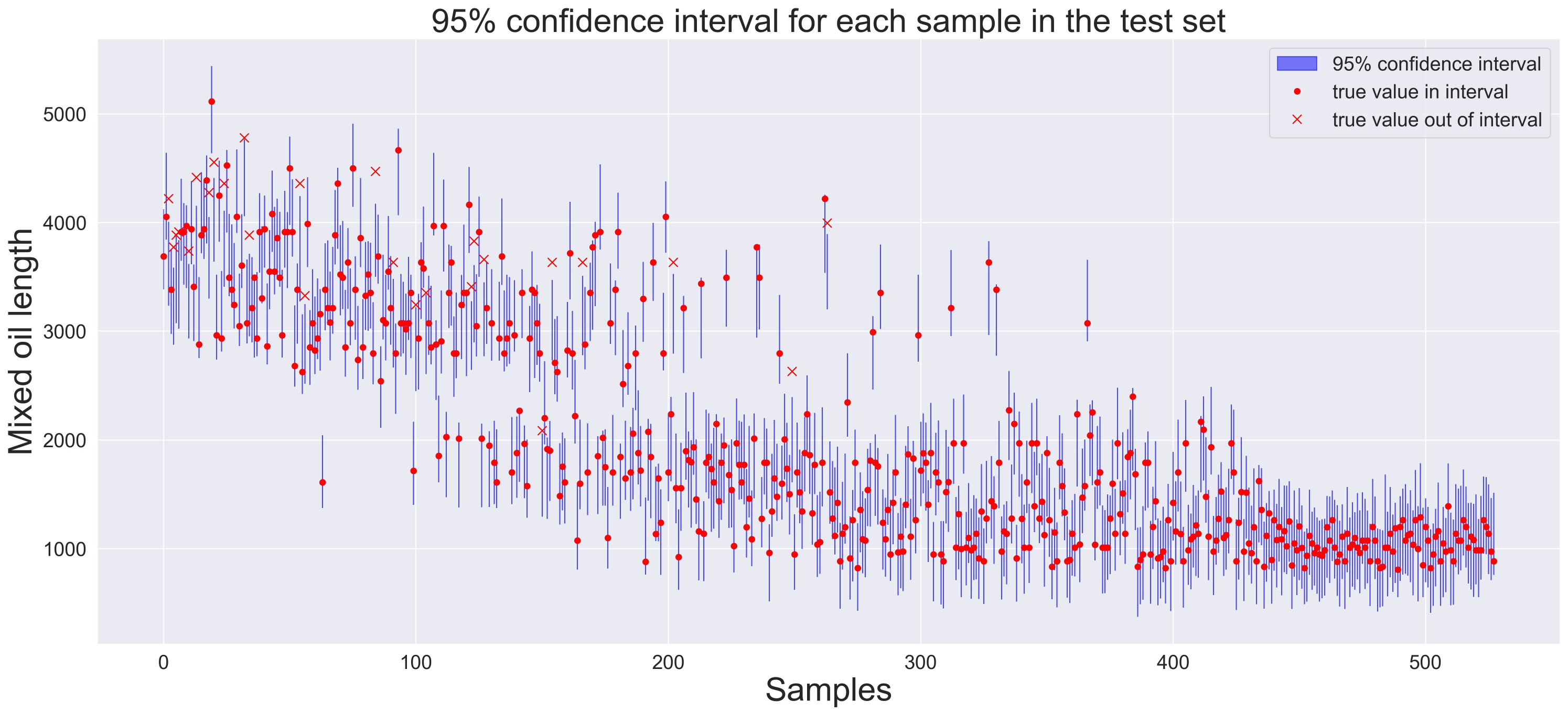}
    \end{minipage}

    \caption{Interval estimations of mixed oil lengths for the third pipeline with confidence levels of 0.9 and 0.95. At a confidence level of 90\%, there are 36 samples falling outside the intervals, achieving a coverage rate of 93.2\%, with an average radius of 317.0 for the intervals.
At a 95\% confidence level, there exist 26 samples outside the intervals, exhibiting a coverage rate of 95.1\%, with the intervals having an average radius of 378.5.
%With a confidence level of 0.9, there are 36 samples outside the intervals, with a coverage rate of 0.931 and an average radius of 317.0 for the intervals. At a confidence level of 0.95, there are 26 samples outside the intervals, with a coverage rate of 0.951 and an average radius of 378.5 for the intervals.}%All parameters are mentioned in Section \ref{Parameter tuning and modeling process} and Section \ref{Interval estimation of the length of the mixed oil}.
}
    \label{fig:confidence interval}
 \end{figure*}

We  randomly select $70\%$ samples of the first two pipelines as the training set and $30\%$ as the validation set. Samples from the training set are used to train our models and samples from the validation set are used to tune parameters. 
The samples from the third pipeline are taken as the test set. % to evaluate the generalization ability of our proposed model. % in Section \ref{Results comparison}.
For each feature $X$ from the test set, we draw $N=200$ pseudo samples from the diffusion model. 
The generated confidence interval estimations are shown in Figure \ref{fig:confidence interval} with confidence levels 90\% and 95\%. .

Besides our proposed method, we also consider the Austin-Palfrey equation, the Gaussian mixture regression \cite{yuan2023knowledge}, Chen's method \cite{chen2021novel}, and Yuan's method \cite{yuan2021robust}. From Table \ref{tab:prob}, we can find that the Austin-Palfrey formula is highly likely to significantly underestimate the length of the oil mix, with a probability exceeding 98\%, while the approaches in \cite{yuan2023knowledge}, \cite{chen2021novel} and \cite{yuan2021robust} exhibit probabilities of 88.8\%, 35.7\% and 47.4\%, respectively.
Our interval estimation method effectively provides a confidence upper limit for the mixed oil length at the confidence level $\alpha$. 
It is worth noting that, at the expense of excluding approximately 300 more meters of oil, the proposed confidence upper limit underestimates the length of mixed oil with probabilities of 6.8\% and 4.9\% when $\alpha$ is set to 90\% and 95\%, respectively.

Therefore, pipeline companies have the flexibility to determine the confidence level $\alpha$ based on their specific requirements and, accordingly, can exclude the mixed oil at the upper limit.

\begin{figure*}[thbp!]
    \centering

    \begin{minipage}[t]{1.0\linewidth}
        \centering
        \includegraphics[width=0.8\linewidth]{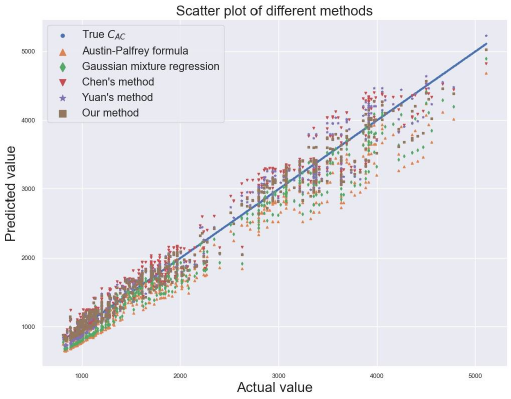}
    \end{minipage}
    
    \caption{Scatter plot comparisons among various models for the third pipelines. }
    \label{fig:densities and scatter plots}
 \end{figure*}

Moreover, we use the mean of the pseudo samples generated by the diffusion model as an estimator for mixed oil length. We then compare the prediction accuracy among different methods. Specifically, following \cite{chen2021novel} and \cite{yuan2021robust}, we employ RMSE, R$^2$, and MAE to assess the performance:
\begin{equation}
\label{rmse}
\begin{split}
    \mbox{RMSE}=\sqrt{\dfrac{1}{n}\sum_{i=1}^{n}( y_i-\Hat{y}_i)^2},
\end{split}
\end{equation}
\begin{equation*}
\label{r2}
\begin{split}
    R^2=1-\dfrac{\sum_{i=1}^{n}( y_i-\Hat{y}_i)^2}{\sum_{i=1}^{n}( y_i-\Bar{y}_i)^2},
\end{split}
\end{equation*}
\begin{equation*}
\label{mae}
\begin{split}
   \mbox{MAE}=\dfrac{1}{n}\sum_{i=1}^{n}| y_i-\Hat{y}_i|,
\end{split}
\end{equation*}
where $y_i$, $\Hat{y}_i$, and $\Bar{y}_i$ represent the actual value, predicted value, and sample mean of the length of the mixed oil in the test set. A large $R^{2}$ is preferred, while small RMSE and MAE are desired. The results are shown in Table \ref{tab:results}. It is observed that our proposed method outperforms the other four methods in terms of the three criteria. For example, in comparison to Yuan's method, which serves as the best baseline, our approach achieves a reduction in MAE by approximately 14\%. Furthermore, Figure \ref{fig:densities and scatter plots} compares the scatter plots of predicted values from various methods with the true values for the third pipeline. The closer the scatter plot of predicted values is to that of the true $C_{AC}$, the better the model fitting. It is evident that our proposed method outperforms the other four methods in this comparison.

\begin{table}[thbp!]
\centering
\caption{The probability of underestimating the length of mixed oil  using different methods.}
\begin{tabular}{cc}
\hline
Models & Probability \\
\hline
Austin-Palfrey equation & 98.3$\%$ \\
%\hline
Gaussian mixture regression & 88.8$\%$ \\
%\hline
Chen's method & 35.7$\%$ \\
%\hline
Yuan's method & 47.4$\%$ \\
%\hline
%Our model & 44.7$\%$ \\
%\hline
Upper limit ($\alpha=0.90$) & 6.8$\%$ \\
%\hline
Upper limit ($\alpha=0.95$) & \textbf{4.9}$\%$ \\
\hline
\end{tabular}
% A smaller probability indicates that the model is more useful in practical operation. The best results are in bold font.}
\label{tab:prob}
\end{table}

\begin{table}[thbp!]
 \centering
 \caption{The RMSEs, $R^2$s and MAEs  for predictions of mixed oil length using various methods. }%The content in brackets represents the fitting target of the models. For example, $(C_{AC}-C_{AP})$ indicates the model is fitting $C_{AC}-C_{AP}$ and $(C_{AC})$ means fitting $C_{AC}$. Interval mean refers to taking the mean of interval estimation generated by diffusion model as a point prediction. The best results are in bold font.}
\begin{tabular}{cccc}
\hline
Models & $RMSE$ & $R^2$ & $MAE$\\
\hline%\Xhline{1.5pt}
Austin-Palfrey equation  & 311.065 & 0.916 & 245.037\\
%\hline
Gaussian mixture regression & 244.938 & 0.948 & 177.783 \\ %\hline
Chen's method & 207.007 & 0.963 & 160.773\\
%\hline
Yuan's method  & 185.849 & 0.970 & 144.543\\
%\hline
%Our model ($C_{AC}-C_{AP}$) & \textbf{169.173} & \textbf{0.975} & 127.192\\
%\hline
Our method & \textbf{171.913} & \textbf{0.974} & \textbf{123.991}\\

\hline

\end{tabular}
\label{tab:results}
\end{table}

\section{Conclusion}
\label{Conclusion}

Mixed oil is inevitably formed during the batch transportation in product pipelines. The presence of mixed oil can result in a deterioration of the quality of oil products, impacting factors such as octane number, flash point, and final boiling point. This can lead to the refined products being of substandard quality \cite{du2023deeppipe}. For example, the South China Multi-Product Pipeline Network Company expends millions of CNY annually on the treatment of mixed oil \cite{du2023intelligent}. The utilization of off-specification oil products in machinery can result in equipment malfunctions, potentially compromising the operational reliability of gasoline engines and aviation equipment \cite{sunagatullin2019relevant}.  Therefore, it is of utmost importance to prevent the ingress of mixed oil into the refined oil products.

  In this article, we propose the utilization of the conditional diffusion model to establish a confidence interval for the mixed oil length. Our proposed upper limit of the interval ensures a high probability of excluding mixed oil, addressing the limitations of previous methods that solely provide point estimations. The conventional approaches often result in underestimation of the mixed oil length, contributing to significant quality loss with an extremely high probability. Therefore, the upper limit of the constructed confidence interval can serve as a reference for pipeline companies to exclude mixed oil. This approach can help in reducing unnecessary costs while maintaining high oil product quality.   Additionally, using the mean of the pseudo samples generated by the diffusion model as a point estimator of the mixed oil length proves to be more effective than previous commonly used methods.

  To mitigate the influence of the mixed oil segment on the purity of the oil, personnel on-site commonly set aside a specific buffer during oil batch cutting operations \cite{li2017section}. However, this buffer is often decided solely through experience, with notable interference from human factors \cite{harbert2008automation,zheng2023hybrid}. The interval estimation method proposed in this article offers essential data support in establishing the cutting buffer. This is notably crucial for enhancing the operational efficiency of the refined oil distribution network.

\section*{Credit author statement}
Yanfeng Yang: Conceptualization, Methodology, Writing - Original Draft. 
Lihong Zhang:  Writing - Original Draft.
Ziqi Chen: Supervision, Writing - Original draft, Writing - Review \& Editing. 
Miaomiao Yu:  Validation. 
Lei Chen: Data curation.

\section*{Declaration of competing interest }
To the best of our knowledge and belief, this manuscript has neither been published in whole nor in part, nor is it currently under consideration for publication elsewhere. We declare that we have no known competing financial interests or personal relationships that could be perceived to influence the work reported in this paper.

\section*{Data availability }
The data that has been used is confidential.

\section*{Acknowledgement }
Dr. Ziqi Chen’s work was partially supported by the National Key R\&D Program of China
(2021YFA1000100 and 2021YFA1000101), National Natural Science Foundation of China (NSFC)
(12271167, 11871477 and 72331005), and Natural Science Foundation of Shanghai (21ZR1418800).

\bibliographystyle{unsrt}
\bibliography{references}

\appendix
\section{Appendix}
This appendix provides supplementary information to complement the main  paper, including the detailed descriptions of the conditional diffusion model \cite{han2022card} and a comparative analysis with alternative confidence interval estimation methods.

\subsection{Diffusion model}
\label{appendix:diffusion model}
In practical scenarios, the conditional distribution of $Y|X$ is  unknown. We propose to learn this distribution using the diffusion model. Specifically, let $\{x_i, y_i\}_{i=1}^{n}$ represent our training dataset, consisting of $n$ samples used to train the  diffusion model. In this context, $x_i$ denotes the $d_x$-dimensional pipeline features, and $y_i$ represents the corresponding mixed oil length. 
The conventional diffusion model is primarily designed for learning unconditional distributions. We aim to learn the conditional distribution of $Y|X$. In our approach, we integrate a pre-trained model $f_\phi(X)$ into the original diffusion model \cite{han2022card}, enabling it to capture the conditional distribution of the original data given $f_\phi(X)$ (i.e., $Y|f_\phi(X)$). When $f_\phi(X)$ performs well, $Y|f_\phi(X)$ closely approximates $Y|X$. Utilizing this conditional distribution $Y|f_\phi(X)$, we input new pipeline features $X$ into $Y|f_\phi(X)$ to generate pseudo samples. These pseudo samples are  used to construct the confidence interval estimation as described in Section \ref{Confidence interval}.

\begin{figure*}[thbp!]
   \centering
   \begin{minipage}[t]{1.0\linewidth}
       \centering
       \includegraphics[width=1\linewidth]{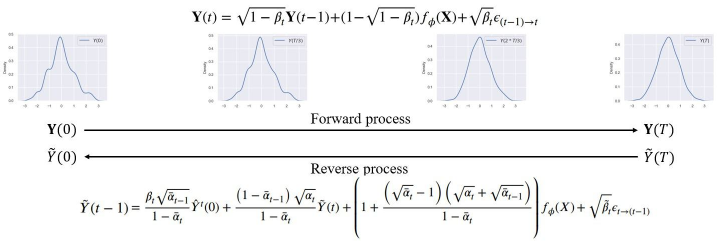}
    \end{minipage}

    \caption{The diagram of diffusion model.}
    \label{fig:structure of Diffusion model}
 \end{figure*}

\begin{table*}[thbp!]
\centering
\caption{The meanings of notations  in the Appendix.}
\begin{adjustbox}{max width=\textwidth, }
\begin{tabular}{cc}
\hline
Notations & Meanings \\
\hline

$\alpha_t$ & $\alpha_t=1-\beta_t$. \\
$\bar{\alpha}_t$ & $\bar{\alpha}_t=\prod_{s=1}^{t}\alpha_s$. \\
$f_{\phi}(X)$ & A pre-trained model, predicting $E(Y|X)$. \\
$\textbf{X}$ & $\textbf{X}=(x_1,x_2,\cdots,x_n)^T$. \\
$\textbf{Y}(0)$ & $\textbf{Y}(0)=(y_1,y_2,\cdots,y_n)^T$. \\
$\epsilon_{i \rightarrow j}$ & A standard $n$-dimensional Gaussian variable, used to transform $\textbf{Y}(i)$ to $\textbf{Y}(j)$. \\
$I_{n \times n}$ & The $n$-dimensional identity matrix. \\
$N(0,I_{n \times n})$ & A standard $n$-dimensional normal distribution. \\
$\textbf{Y}(t)$ & The latent variables in the forward process of the diffusion model. \\
$\textit{\textbf{L}}_{\theta}$ & The loss function of the diffusion model. \\
${nn}_{\theta}$ & A deep neural network, $\theta$ represents the parameters of the deep neural network. \\
$\tilde{Y}(t)$ & The latent variables in the reverse process of the diffusion model. \\
$\Hat{Y}^t(0)$ & An estimation of $Y(0)$ based on $Y(t)$. \\
$\tilde{\mu}\left(Y(0),\tilde{Y}(t),X\right)$ & The mean in (\ref{Y(t-1),Y(t),Y(0),f(X)}).  \\
$\tilde{\beta_t}$ & The variance in  (\ref{Y(t-1),Y(t),Y(0),f(X)}). \\
\hline
\end{tabular}

\label{tab:notation_appendix}
\end{adjustbox}
\end{table*}

Let $t\in \{0,1,\cdots,T\}$ be the time steps in both the forward and reverse processes in the diffusion model. Consider the pre-trained model $f_\phi(X)$, which  is used to predict $Y$ and will be introduced in Section \ref{Parameter tuning and modeling process}. Let $\{\beta_t\}_{t=1}^T$ with $\beta_t\in(0,1)$ be the diffusion schedule. Define $\alpha_t:=1-\beta_t$ and $\bar{\alpha}_t:=\prod_{s=1}^{t}\alpha_s$. Let $\textbf{Y}(t)$, for $t=1,\cdots,T$, represent the latent variables in  the forward  processes. Define $\textbf{X}:=(x_1,x_2,\cdots,x_n)^T$ and $\textbf{Y}(0):=(y_1,y_2,\cdots,y_n)^T$.
For $t=1,\cdots,T$, define \begin{equation*}
\label{forward}
    \textbf{Y}(t)=\sqrt{1-\beta_t} \textbf{Y}(t-1) + \left(1-\sqrt{1-\beta_t}\right)f_\phi(\textbf{X})+\sqrt{\beta_t}\epsilon_{(t-1) \rightarrow t}, 
\end{equation*}

where $\epsilon_{(t-1) \rightarrow t} \sim N(0,I_{n\times n})$ with $I_{n\times n}$ being the $n-$dimensional identity matrix. Thus, we have 
\begin{equation}
\label{Y(t),Y(t-1),f(X)}
    \textbf{Y}(t)|\textbf{Y}(t-1),f_\phi(\textbf{X})  \sim  N\left(\sqrt{\alpha_t} \textbf{Y}(t-1) + \left(1-\sqrt{\alpha_t}\right)f_\phi(\textbf{X}),\beta_t I \right).
\end{equation}
Then, we obtain
\begin{equation*}
\label{re-parameterization trick}
\begin{split}
    \textbf{Y}(t) & =\sqrt{1-\beta_t}  \Big[\sqrt{1-\beta_{t-1}} \textbf{Y}(t-2)\\
    &~~+\left(1-\sqrt{1-\beta_{t-1}}\right)f_\phi(\textbf{X})  +\sqrt{\beta_{t-1}}\epsilon_{(t-2) \rightarrow (t-1)} \Big]\\
    &~~+ \left(1-\sqrt{1-\beta_t}\right)f_\phi(\textbf{X})+\sqrt{\beta_t}\epsilon_{(t-1) \rightarrow t}  \\ & = \sqrt{\alpha_t \alpha_{t-1}} \textbf{Y}(t-2)+\left(1- \sqrt{\alpha_t \alpha_{t-1}} \right)f_\phi(\textbf{X})  \\ &~~+\sqrt{1-\alpha_t \alpha_{t-1}} \epsilon_{(t-2) \rightarrow t},
\end{split}
\end{equation*}
where  $\epsilon_{(t-2) \rightarrow t} \sim N(0,I_{n\times n})$. Similarly, we represent $\textbf{Y}(t)$ as a function of $\textbf{Y}(0)$:
\begin{equation}
\label{Y(t) repersented by Y(0),f_phi(X)}
\begin{split}
    \textbf{Y}(t) & =\sqrt{\bar{\alpha}_t}\textbf{Y}(0)+\left(1-\sqrt{\bar{\alpha}_t}\right) f_\phi(\textbf{X}) +\sqrt{1-\bar{\alpha}_t} \epsilon_{0 \rightarrow t},
\end{split}
\end{equation}
where $\epsilon_{0 \rightarrow t} \sim N(0,I_{n\times n}).$
Thus, the distribution of $\textbf{Y}(t)$ given $\textbf{Y}(0)$ and $f_\phi(\textbf{X})$ can be written as:
\begin{equation}
\label{Y(t)|Y(0),f_phi(X)}
\begin{aligned}
    \textbf{Y}(t)|\textbf{Y}(0),f_\phi(\textbf{X}) \sim N&\left(\sqrt{\bar{\alpha}_t}\textbf{Y}(0)+(1-\sqrt{\bar{\alpha}_t}) f_\phi(\textbf{X}),\right.\\
    &\left.(1-\bar{\alpha}_t) I_{n\times n}\right).
\end{aligned}
\end{equation}
When $T\to \infty$, $\bar{\alpha}_T \to 0$. So, $\textbf{Y}(T)|\textbf{Y}(0),f_\phi(\textbf{X})$ converges to $N\left(f_\phi(\textbf{X}),I_{n\times n}\right)$. % from which we draw noises.

Next, we describe the procedure for generating pseudo samples of $Y|X$ based on a given new pipeline feature $X \in \mathbb{R}^{1 \times d_x}$. In particular, this is achieved through the reverse process of the previously mentioned forward process. Initially, we generate noise $\tilde{Y}(T)$ from $N\left(f_\phi(X),1\right)$. Subsequently, we aim to transform $\tilde{Y}(T)$ into high-quality pseudo samples of $Y|X$, denoted as $\tilde{Y}(0)$.

Let $\tilde{Y}(t),t=T-1,\cdots,1$ be the latent variables in the reverse process. % and they preserve the properties of $Y(t),t=1,\cdotsT$. 
According to Bayes' theorem, (\ref{Y(t),Y(t-1),f(X)}), (\ref{Y(t)|Y(0),f_phi(X)}) and the Markov property of the forward process, we  derive the distribution of $\tilde{Y}(t-1)$ conditional on $\left(\tilde{Y}(t), Y(0), f_\phi(X)\right)$:
\begin{equation}
\label{Y(t-1),Y(t),Y(0),f(X)}
    \tilde{Y}(t-1)|\tilde{Y}(t), Y(0),f_\phi(X)  \sim  N\left(\tilde{\mu}\left(Y(0),\tilde{Y}(t),X\right),\tilde{\beta_t}\right),    
\end{equation}
where
\begin{equation*}
\label{mean of reverse}
\begin{split}
    \tilde{\mu}\left(Y(0),\tilde{Y}(t),X\right)&:=\dfrac{\beta_t \sqrt{\bar{\alpha}_{t-1}}}{1-\bar{\alpha}_{t}}Y(0)+\dfrac{\left(1-\bar{\alpha}_{t-1}\right)\sqrt{\alpha_t}}{1-\bar{\alpha}_{t}}\tilde{Y}(t)+ \\ &\left( 1+\dfrac{\left(\sqrt{\bar{\alpha}_{t}}-1\right)\left(\sqrt{\alpha_t}+\sqrt{\bar{\alpha}_{t-1}}\right)}{1-\Bar{\alpha}_t}\right) f_\phi(X)
\end{split}
\end{equation*}
and
\begin{equation}
\label{beta t}
    \tilde{\beta_t}:=\dfrac{1-\Bar{\alpha}_{t-1}}{1-\Bar{\alpha}_{t}}\beta_t.
\end{equation}
%As long as we know parameters $\tilde{\mu}$ and $\tilde{\beta_t}$, by (\ref{Y(t-1),Y(t),Y(0),f(X)}), we  generate $\textbf{Y}(t-1)$ through $\textbf{Y}(t)$, $\textbf{Y}(0)$, and $f_\phi(\textbf{X})$ using:
$\bar{\alpha}_t$, $\alpha_t$ and $\tilde{\beta_t}$ can be directly calculated because $\beta_t$'s are constants predetermined by us. Consequently, we can generate $\tilde{Y}(t-1)$ using (\ref{Y(t-1),Y(t),Y(0),f(X)}) by incorporating $\Tilde{Y}(t)$, $Y(0)$, and $f_\phi(X)$. The generation process is expressed as follows:
\begin{equation}
\label{infer Y(t-1),easy version}
    \Tilde{Y}(t-1)=\tilde{\mu}\left(Y(0),\Tilde{Y}(t),X\right)+\sqrt{\tilde{\beta_t}}\epsilon_{t \rightarrow (t-1)},  
\end{equation}
where $\epsilon_{t \rightarrow (t-1)} \sim N(0,1)$, for $t=T,\cdots,1$. By repeating this process $N$ times, we generate $N$ pseudo samples. 

However, in the reverse process, we do not possess  $Y(0)$. So, we use the predicted $Y(0)$ instead of directly using $Y(0)$. Specifically,  by  (\ref{Y(t) repersented by Y(0),f_phi(X)}),
\begin{equation*}
\label{infer Y(0)}
    \textbf{Y}(0)=\dfrac{1}{\sqrt{\bar{\alpha}_t}}\left(\textbf{Y}(t)-\left(1-\sqrt{\bar{\alpha}_t}\right)f_\phi(\textbf{X})-\sqrt{1-\bar{\alpha}_t}\epsilon_{0 \rightarrow t}\right).
\end{equation*}

We  use a deep neural network ${nn}_{\theta}$ to approximate $\epsilon_{0 \rightarrow t}$. The loss function can be formulated as:
\begin{equation*}
\label{loss function}
    \textit{L}=\mathbb{E}_{t, \textbf{Y}(0) , \textbf{Y}(t)} \| \epsilon_{0 \rightarrow t} - {nn}_{\theta}\left(\textbf{X},\textbf{Y}(t), f_\phi(\textbf{X}), t\right)\|^2.
\end{equation*}
With the loss function, we can leverage optimization algorithms like Adam \cite{kingma2014adam} to train the neural network ${nn}_{\theta}$, aiming to bring ${nn}_{\theta}$ sufficiently close to $\epsilon_{0 \rightarrow t}$. Then, in the reverse process (\ref{infer Y(t-1),easy version}),  $Y(0)$ is replaced by its predicted value:
\begin{equation*}
\label{recursive formula}
\begin{split}
    \Hat{Y}^t(0)= \dfrac{1}{\sqrt{\bar{\alpha}_t}}\Big[ & \Tilde{Y}(t)-\left(1-\sqrt{\bar{\alpha}_t}\right)f_\phi(X)\\ & -\sqrt{1-\bar{\alpha}_t} nn_{\theta}\left(X,\Tilde{Y}(t), f_\phi(X), t \right)\Big]. 
\end{split}
\end{equation*}

Refer to Figure \ref{fig:structure of Diffusion model} and Algorithm \ref{algo:card} for the detailed process and algorithm that illustrates the conditional diffusion model used to generate pseudo samples of $Y|X$.

\subsection{Comparative experiment}

\begin{table*}[thbp!]
\centering
\caption{The coverage rates and average lengths of different confidence intervals with confidence levels of 90\% and 95\%.}
\begin{adjustbox}{max width=\textwidth, }
\begin{tabular}{ccccc}
\hline
Methods & Coverage rate (90\%) & Average length (90\%) & Coverage rate (95\%) & Average length (95\%)\\
\hline
Linear quantile regression & 95.1\% & 413.0 & 96.4\% & 452.4 \\
Random forest quantile regression & 80.0\% & 331.7 & 85.8\% & 381.9 \\
Conformalized quantile regression & 79.2\% & 251.2 & 84.3\% & 291.8 \\
Ours & 93.2\% & 317.0 & 95.1\% & 387.5 \\
\hline
\end{tabular}
% A smaller probability indicates that the model is more useful in practical operation. The best results are in bold font.}
\label{tab:comparison}
\end{adjustbox}
\end{table*}

\begin{figure*}[htbp]   
    \centering
    \begin{minipage}{1.0\linewidth}
        \centering
        \includegraphics[width=0.9\linewidth]{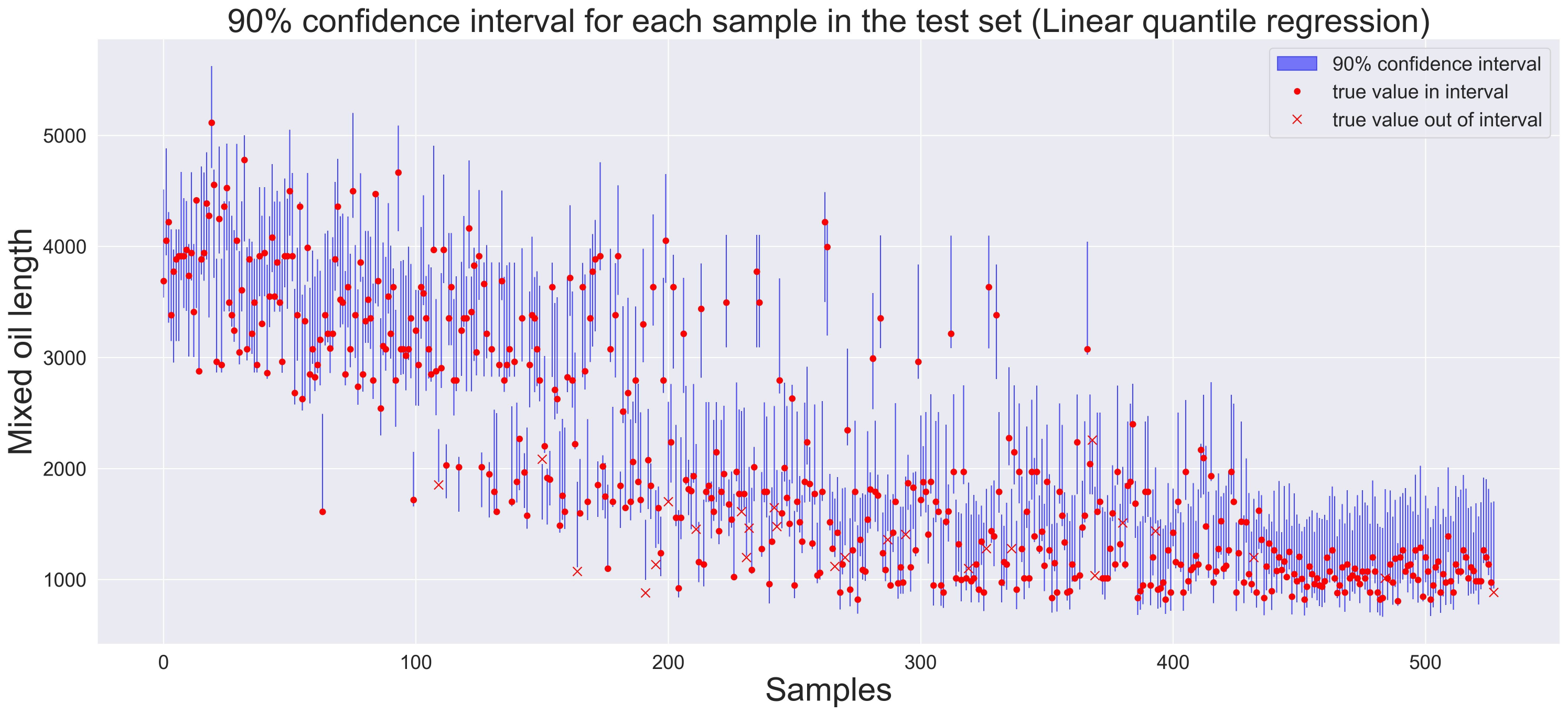} 
        \\ \vspace{10mm}
        \includegraphics[width=0.9\linewidth]{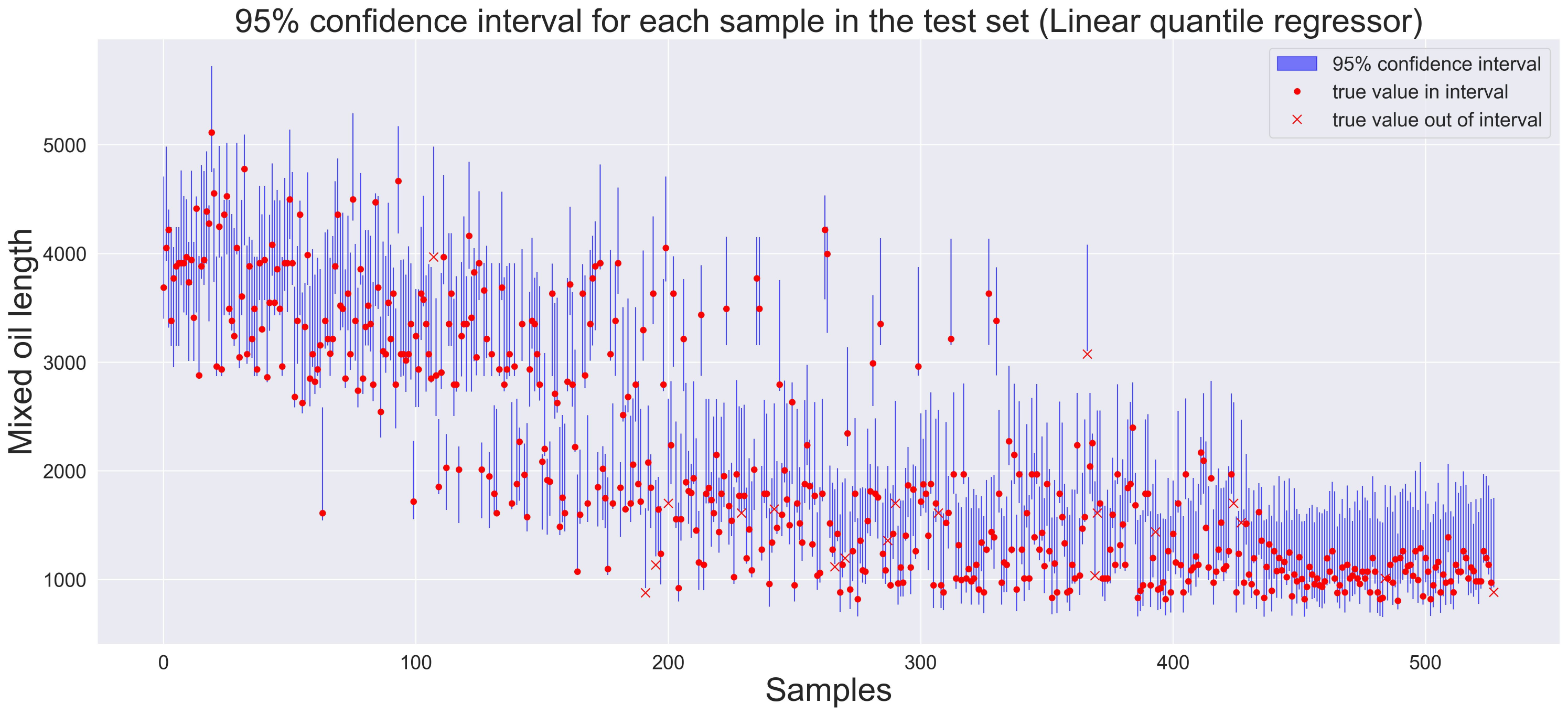}
    \end{minipage}
  \caption{Interval estimations constructed by the linear quantile regression with confidence levels of 0.9 and 0.95. Coverage rate and average length of the intervals can be found in Table \ref{tab:comparison}.} 
  \label{fig:comparison 1}       
\end{figure*}

\begin{figure*}[htbp]   
    \centering
    \begin{minipage}{1.0\linewidth}
        \centering
        \includegraphics[width=0.9\linewidth]{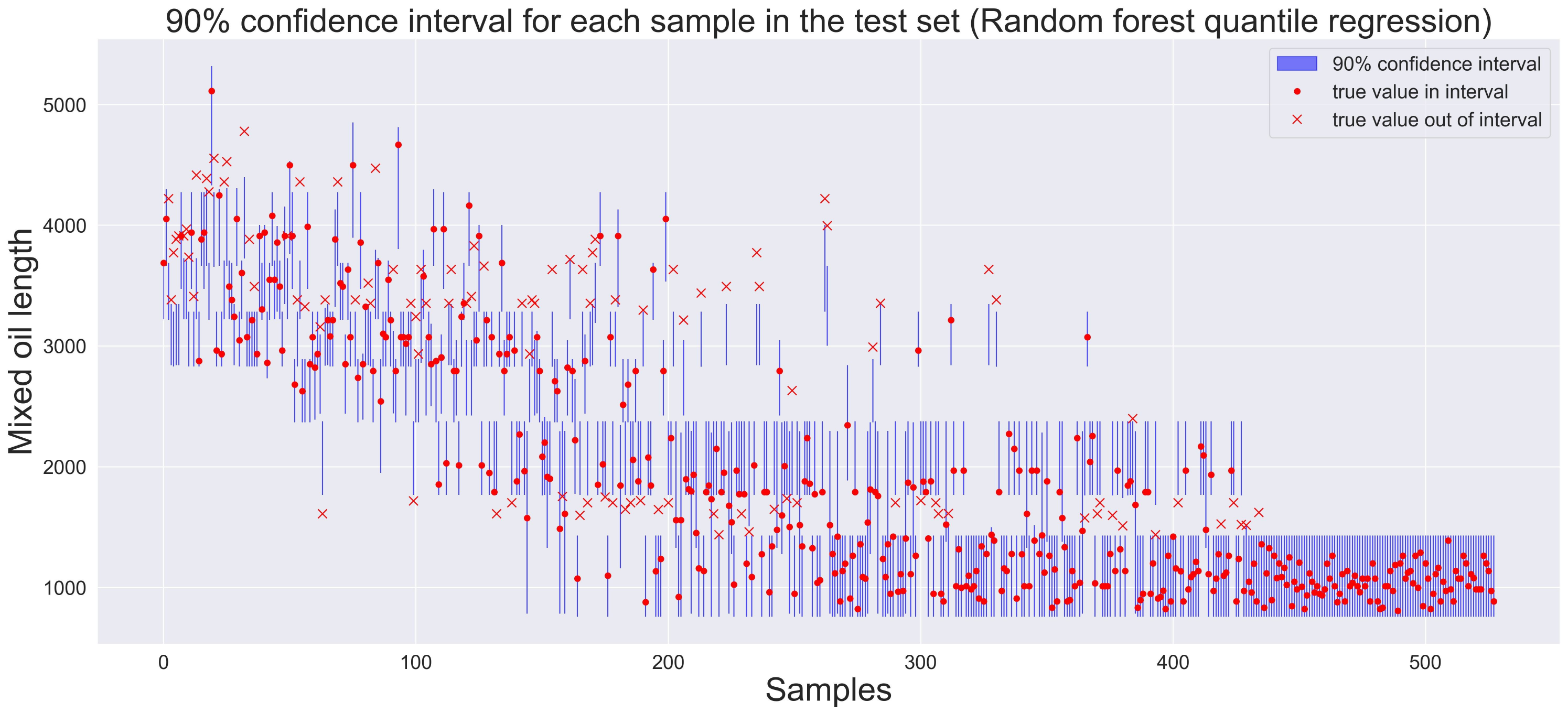}
        \\ \vspace{10mm}
        \includegraphics[width=0.9\linewidth]{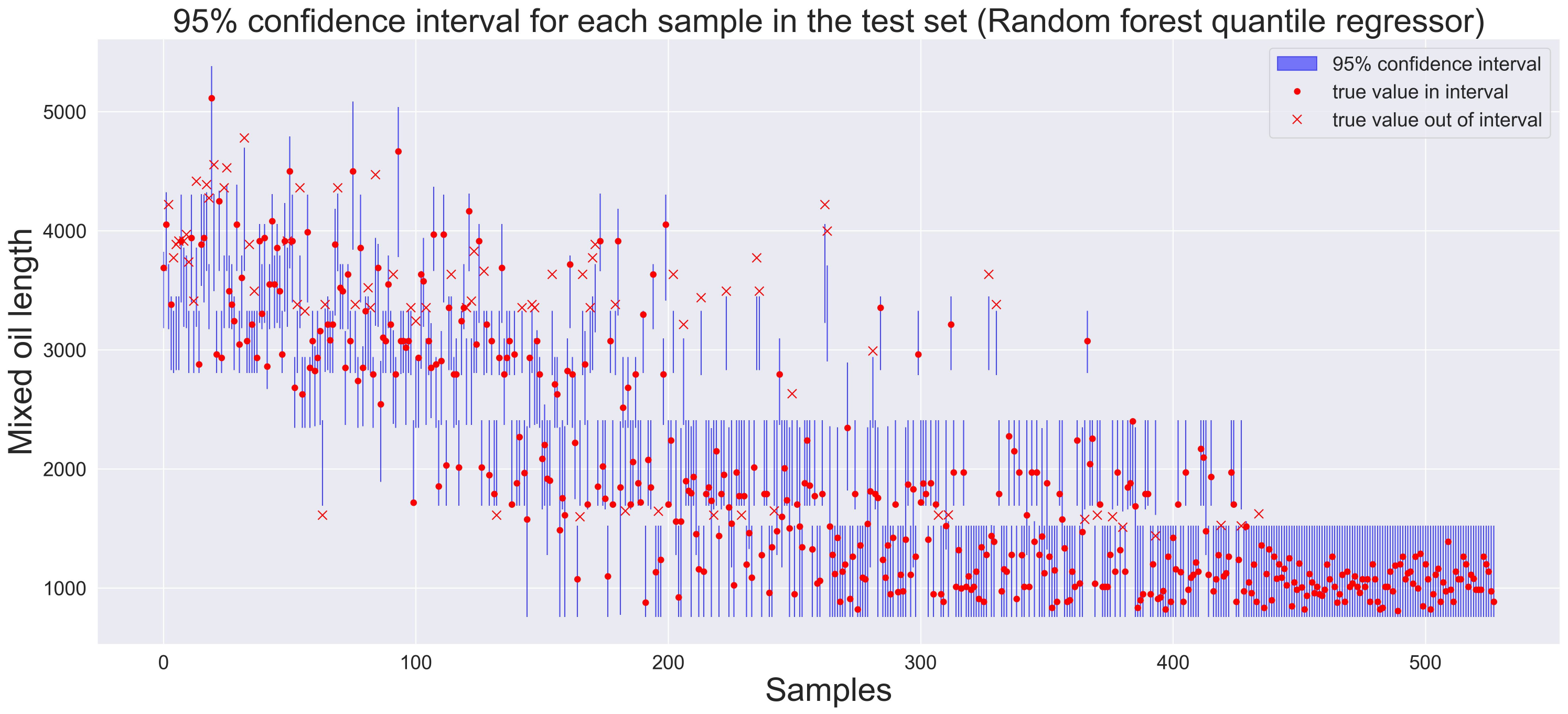}
    \end{minipage}
  \caption{Interval estimations constructed by the random forset quantile regression with confidence levels of 0.9 and 0.95. Coverage rate and average length of the intervals can be found in Table \ref{tab:comparison}.} 
  \label{fig:comparison 2}       
\end{figure*}

\begin{figure*}[htbp]   
    \centering
    \begin{minipage}{1.0\linewidth}
        \centering
        \includegraphics[width=0.9\linewidth]{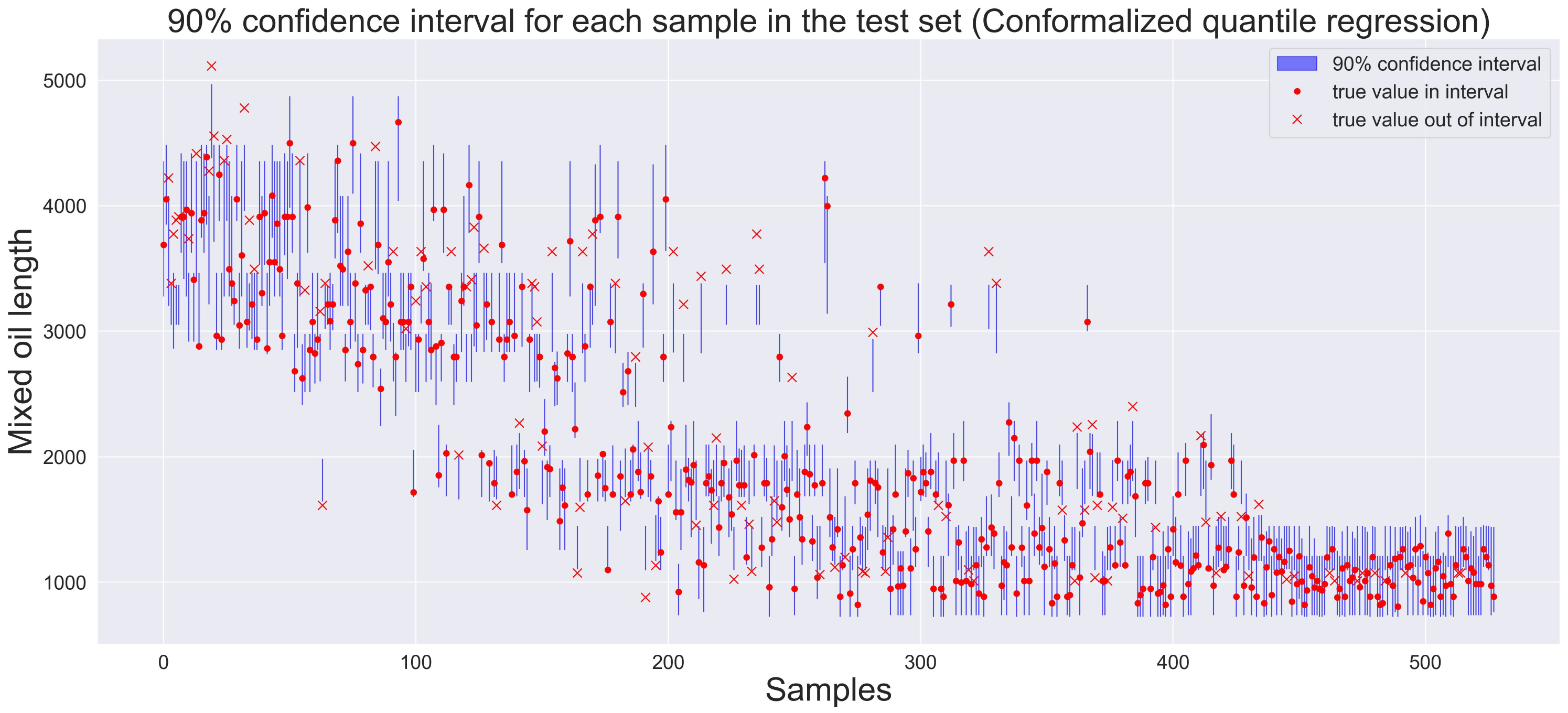}
        \\ \vspace{10mm}
        \includegraphics[width=0.9\linewidth]{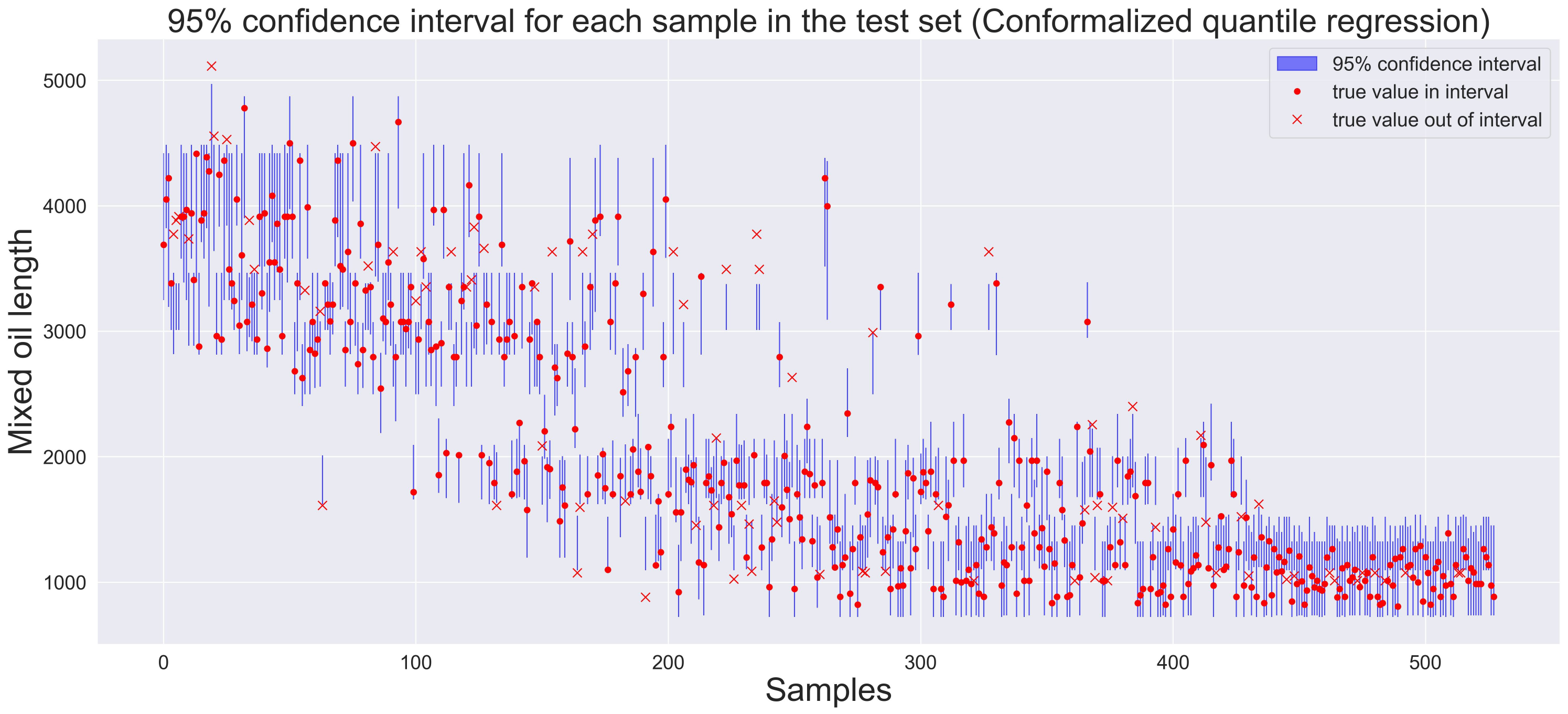}
    \end{minipage}
  \caption{Interval estimations constructed by the conformalized quantile regression with confidence levels of 0.9 and 0.95. Coverage rate and average length of the intervals can be found in Table \ref{tab:comparison}.} 
  \label{fig:comparison 3}       
\end{figure*}

The proposed method constructs confidence intervals by directly learning the conditional distribution of $Y|X$. Besides, other commonly used methods for constructing confidence intervals include the quantile regression and the conformal prediction. So, we use the linear quantile regression \cite{hao2007quantile}, the random forest quantile regression \cite{JMLR:v7:meinshausen06a}, and the conformalized quantile regression \cite{romano2019conformalized} to construct confidence interval estimations, and compare the results with ours. Detailed results can be found in the Table \ref{tab:comparison},  Figures \ref{fig:comparison 1},  \ref{fig:comparison 2} and \ref{fig:comparison 3}.

Table \ref{tab:comparison} reveals that the linear quantile regression produces relatively longer interval estimations, potentially leading to resource wastage. The average length of interval estimation provided by the random forest quantile regression is similar to ours, but this method fails to cover enough samples. The conformalized quantile regression yields notably short interval estimations, but its coverage rate is the lowest. Considering both the coverage rate and the average length, our approach has the most favorable performances.

\end{document}